\documentclass[showpacs,showkeys,amsmath,amssymb,preprint,nofootinbib]{revtex4}
\usepackage{graphicx}
\usepackage{epsfig}
\usepackage{bm}
\usepackage{float}
\usepackage[T1]{fontenc}
\usepackage[utf8]{inputenc}
\usepackage{graphicx}
\usepackage{bm}
\usepackage{amsmath}
\usepackage{xcolor}
\DeclareUnicodeCharacter{2212}{\textendash}
\def\beq{\begin{eqnarray}}

\def\eeq{\end{eqnarray}}

\begin{document}

\title[DEA]{Phantom cosmologies from QCD ghost dark energy}

\author{Miguel Cruz$^{a,}$\footnote{miguelcruz02@uv.mx}, Samuel Lepe$^{b,}$\footnote{samuel.lepe@pucv.cl} and Germ\'an E. Soto$^{b,}$\footnote{german.soto.z@mail.pucv.cl}}

\affiliation{$^{a}$Facultad de F\'{\i}sica, Universidad Veracruzana 91097, Xalapa, Veracruz, M\'exico \\
$^{b}$Instituto de F\'\i sica, Facultad de Ciencias, Pontificia Universidad Cat\'olica de Valpara\'\i so, Av. Brasil 2950, Valpara\'\i so, Chile}

\date{\today}

\begin{abstract}
We review a dynamical dark energy model scarcely studied in the literature and we introduce two possible generalizations. We discuss separately the behavior of the original model and a minimal extension of it by exploring some early and late times limits, we find that the cosmic components are related by their parameters state. In order to have access to the phantom regime we present two dark energy densities inspired from the holographic approach and from the emergent entropic forces model in the early universe. For the first case we obtain a Type III singularity and in the second proposal we have a transition from decelerated to accelerated cosmic expansion that starts as phantom type. However, the final stage of the universe is a de Sitter state.
\end{abstract}



\maketitle

\section{Introduction}
\label{sec:intro}
Several proposals for the unknown form of energy density in the universe or simply dark energy can be found in the literature. The aim of such models is to seek alternatives to the pervasive philosophy of the cosmological constant (misnamed dark energy). Despite the $\Lambda$CDM success to fit data, the updated observational evidence seem to indicate some tensions and issues in the model, see for instance Ref. \cite{valentino} and references therein. In addition, it is difficult to explain the possibility of crossing the phantom barrier ($\omega =-1$), a barrier whose temperature is zero in the cosmic fluid analogy, if we consider the cosmological constant approach. The possibility that we are living in a phantom stage ($\omega <-1$) it is not ruled out by the current observational data \cite{planck}, then the study of cosmological models allowing the phantom scenario becomes relevant.

The origin for the cosmological constant can be solved by assuming the existence of vacuum energy, such vacuum energy is produced from vacuum fluctuations and this mechanism is well understood in the context of quantum field theory. However, the tiny value obtained for the energy density of the cosmological constant is a severe problem and was widely discussed in \cite{weinberg}. This gave rise to plethora of works, as can be seen in the literature. For instance, in Ref. \cite{arkani} the value of the present time energy density for the cosmological constant is related to physics at electroweak scale. 

The observed cosmic expansion requires the right magnitude for the cosmological constant, in Ref. \cite{ghost1} we can find an interesting proposal: gravity is promoted to be a low energy effective interaction, then the cosmological constant emerges from the contribution of some ghost fields which are supposed to be present in the low energy effective theory of QCD. On the other hand, ghost fields are required to exist for the resolution of the $U(1)$ problem, these auxiliary fields are not physical propagating degrees of freedom and in order to avoid the appearance of new degrees of freedom they must be decoupled from the physical sector; see Ref. \cite{ghost2} for a complete review in this topic. However, it was found that when the ghost fields are in a curved spacetime, they contribute to the vacuum energy with a small amount, $\rho \propto H \Lambda^{3}_{QCD}$, where $H$ is the Hubble parameter and $\Lambda_{QCD}$ is the QCD mass scale, this is usually termed as Veneziano ghost model. This latter result provides the right magnitude for the dark energy that explains the current accelerated expansion. As can be seen, this running contribution to vacuum energy departs from a constant behavior but the dynamical scenario for dark energy is also an interesting approach to explore \cite{jje}. See also Ref. \cite{ghost3}, where a dark energy proportional to the Hubble parameter is also obtained from a generalized form of the QCD ghost model. A very interesting characteristic of a vacuum energy linear in the Hubble parameter is the de Sitter phase that can be obtained when such energy is introduced in the Friedmann constraint. However, the origin of this kind of running vacuum energy can be explained formally from the study of dynamics of the topologically nontrivial sectors in a strongly coupled QCD-like gauge theory in an expanding universe with a given hyperbolic space as background and nontrivial holonomy. In this case the energy is due to the tunneling transitions between different topological sectors and in consequence can not be formulated in terms of any local propagating degrees of freedom, therefore the non trivial contributions to the vacuum energy can not be renormalized by any UV terms. See Refs. \cite{bar1, bar2}\footnote{and references therein for a contextualization on the origin of running vacuum energy.}, where the aforementioned formalism was proposed and profoundly discussed. Then the ghost fields description for the running vacuum energy linear in $H$ can be bypassed. It is worthy to mention that the linear correction for the vacuum energy obtained in \cite{bar1, bar2} is consistent with lattice simulations, see for instance Ref. \cite{lattice1}, where it is studied the dependence of vacuum energy on the size of the system and Ref. \cite{lattice2} where the rate particle production turns out to be linearly proportional to the Hubble constant, rather than $H^{2}$ for a de Sitter background.\\ 

In Refs. \cite{ansatz1} and \cite{ansatz2} we can find two proposals for dark energy dubbed as {\it ghost dark energy} since they preserve the dependence on the Hubble parameter in their energy densities; in both cases a transition from decelerated to accelerated expansion takes place at the past and for the  late times behavior of the universe a de Sitter evolution is obtained, i.e., the transition to a phantom scenario is not allowed. According to latest astrophysical data, some extensions of the QCD ghost dark energy model are viable to describe the late times behavior of the observable universe; see for instance Refs. \cite{refe1, refe2}, where was found that this kind of dark energy model could exhibit (in some cases) a better fit of the cosmological data sets than the $\Lambda$CDM model. Besides, in Ref. \cite{refe3} an improvement in the fit of cosmological data with respect to the isotropic case was obtained with the inclusion of anisotropy effects in the QCD ghost model. However, the parameter state associated to the best fit values of the cosmological parameters in the aforementioned references still corresponds in all cases studied to a quintessence fluid at present time. Then this kind of dark energy model must be explored in more detail and extended to the context of phantom cosmologies.

The aim of this work is to establish some extensions of these ghost dark energy models in order to have access to the phantom regime. As we will discuss below, one of the models admits a Type III singularity and our second model exhibits a transient phantom scenario which tends to a de Sitter evolution. Due to the fact that not all physical quantities diverge during the phantom regime, we consider that both scenarios deserve deeper exploration. In the present work we will focus on the behavior at early and late times of the discussed dark energy models.

This paper is organized as follows: In Section \ref{sec:ii} we review a ghost dark energy proposal, highlighting its early and late times limits. In Section \ref{sec:iii} we discuss a first extension of the aforementioned model and its behavior as a function of the cosmic time. In Section \ref{sec:iv}, we explore the phantom scenario arising from an extension of the original ghost dark energy model which was inspired from the holographic approach for dark energy. In Section \ref{sec:v} we study some general properties of a model which describes a phantom evolution with no singularities, except for the initial state of the parameter state. Section \ref{sec:vi} is devoted to our final comments. In this work we will consider $8\pi G=c=k_{B}=1$ units.   
 
\section{Ghost dark energy}
\label{sec:ii}
For this dark energy model the energy density is given by $\rho _{de}=3\alpha H$ \cite{ansatz1}. We restrict ourselves to a flat Friedmann-Lemaitre-Robertson-Walker (FLRW) background, therefore the Friedmann constraint for a cosmic fluid characterized by its energy density $\rho$ and pressure $p$, reads
\begin{equation}
3H^{2}=\rho + \rho_{de}=\rho +3\alpha H, \label{eq:fried1}
\end{equation}
we denote the dark matter sector by $\rho$. Besides, the acceleration equation has the usual form
\begin{equation}
2\dot{H}+3H^{2} = -p-p_{de}. \label{eq:accel}    
\end{equation}
As usual, $H:=\dot{a}/a$ is the Hubble parameter where $a$ is the cosmic scale factor and the dot stands for derivatives w.r.t. cosmic time. From equation (\ref{eq:fried1}) the following expanding solution for the Hubble parameter can be penned straightforwardly
\begin{equation}
    H = \frac{1}{2}\alpha \left( 1+\sqrt{1+\frac{4\rho }{3\alpha ^{2}}}\right). \label{eq:fried2}
\end{equation}
Notice that $\rho=0$ in the above expression leads to the self-accelerated solution $H=\alpha$, or in other words; a constant expansion rate for the cosmic evolution can be obtained as consequence of the dilution of the energy density of the dark matter sector. This self-accelerated solution does not carry instabilities as the one found in the Dvali-Gabadadze-Porrati model, which in turn results inappropriate for doing cosmology \cite{hinterbichler}. Let us consider the case in which both cosmic components do not interact, therefore we can write their corresponding conservation equations as follows
\begin{align}
& \dot{\rho}_{de}+3H\left( 1+\omega _{de}\right) \rho _{de}=0, \label{eq:cons1}\\
& \dot{\rho}+3H\left( 1+\omega \right) \rho =0, \label{eq:cons2}
\end{align}
where we have adopted a barotropic equation of state for each specie, $p_{i} = \omega_{i}\rho_{i}$, the subscript $i$ accounts for different species and $\omega$ is the parameter state. By means of the dark energy Ansatz mentioned previously and Eq. (\ref{eq:cons1}), we can identify the parameter state of dark energy
\begin{equation}
\omega_{de} = -1+\frac{1}{3}\left( 1+q\right), 
\end{equation}%
with the usual definition for the deceleration parameter, $q=-1-\dot{H}/H^{2}$. Alternatively, from the acceleration equation (\ref{eq:accel}) and the above expression for $\omega_{de}$, the deceleration parameter can be written as 
\begin{equation}
q=-1 + \frac{3\left(1+\omega \right)}{2+\rho _{de}/\rho }, \label{eq:alter}
\end{equation}%
thus we can relate the parameters state of both cosmic components as follows
\begin{equation}
\omega_{de}=-1 + \frac{\left( 1+\omega \right)}{2+\rho _{de}/\rho }. \label{eq:omega}
\end{equation}
According to Eq. (\ref{eq:omega}), we establish the following limits based on the behavior of the quotient between energy densities
\begin{align}
& \mbox{early times limit} :\frac{\rho _{de}}{\rho } \ll 1 \ \Longrightarrow \omega_{de}\approx -\frac{1}{2}\left( 1-\omega \right) ,\\
& \mbox{late times limit} :\frac{\rho _{de}}{\rho } \gg 1 \ \Longrightarrow \omega_{de}\approx -1+\left( 1+\omega \right) \frac{\rho }{\rho _{de}} \approx -1,
\end{align}
from the early times limit we can identify the following cases of interest for each parameter state
\begin{eqnarray}
\omega &=&1 \Longrightarrow \omega_{de} \approx 0, \label{eq:stiff}\\
\omega &=&\frac{1}{3} \Longrightarrow \omega_{de} \approx -\frac{1}{3}, \label{eq:rad}\\
\omega &=&0 \Longrightarrow \omega_{de} \approx -\frac{1}{2}. \label{eq:cdm}
\end{eqnarray}%
Thus, in the era of stiff matter dominance given in equation (\ref{eq:stiff}), $\rho_{de}$ behaves as cold dark matter. In the epoch of radiation dominance (\ref{eq:rad}) we have a interesting behavior for the dark energy sector. In the single fluid description of standard cosmology, the case $\omega=-1/3$ describes a Dirac-Milne universe for which, $H(t) \propto t^{-1}$; this kind of universe also emerges uniquely from kinematic relativity and cosmological principle considerations. An interesting work on Milne model is given in Ref. \cite{milne}. In the era dominated by cold dark matter (\ref{eq:cdm}), $\rho_{de}$ behaves as a quintessence fluid with an evolution towards a de Sitter stage. 

Assuming the usual form of the redshift parameter in terms of the scale factor, $1+z=a_{0}/a$, the solution for $\rho \left( z\right)$ obtained from (\ref{eq:cons2}) is given as, $\rho \left( z\right) = \rho \left( 0\right) \left( 1+z\right) ^{3\left( 1+\omega \right)}$. Using this result for dark matter together with Eq. (\ref{eq:fried2}), the parameter state for dark energy given in (\ref{eq:omega}) reads
\begin{equation}
\omega _{de}\left( z\right) =-1+\left( 1+\omega \right) \left[2+\frac{\eta 
}{\left( 1+z\right) ^{3\left( 1+\omega \right) }}\left( 1+\sqrt{1+\frac{2}{\eta}\left( 1+z\right) ^{3\left( 1+\omega \right) }}\right) \right] ^{-1}, 
\end{equation}
where we have defined the positive constant, $\eta :=3\alpha ^{2}/2\rho \left( 0\right)$. From the above expression we can observe that at present time the dark energy sector behaves as quintessence fluid, $\omega_{de}(z=0) = -1 + (1+\omega)/[2+\eta (1+ \sqrt{1+2/\eta})]$ and for the far future, $\omega_{de}(z\rightarrow -1) \rightarrow -1$. Using the Friedmann constraint and evaluating at present time, we can fix the value of $\eta $ according to the normalization condition
\begin{equation}
3H^{2}\left( 0\right) =\rho \left( 0\right) +3\alpha H\left( 0\right)
\rightarrow 1=\Omega \left( 0\right) +\Omega _{de}\left( 0\right),
\end{equation}%
where the fractional energy densities are: $\Omega \left( 0\right) =\rho \left( 0\right) /3H^{2}\left(0\right)$ and $\Omega _{de}\left( 0\right) =\alpha /H\left( 0\right)$. Then, $\eta =\Omega _{de}^{2}\left( 0\right) /2\Omega \left( 0\right)$.

\subsection{Minimal extension of ghost dark energy}
\label{sec:iii}
In this section we explore the dark energy density given as follows \cite{ansatz2}
\begin{equation}
\rho_{de}=3\left( \alpha H+\beta H^{2}\right),
\label{eq:ans2}
\end{equation}%
as can be seen, this Ansatz includes the next leading order term in $H$ of the Veneziano ghost field discussed in the previous section. According to the Friedmann constraint, we can write for all constituents of the universe
\begin{equation}
3H^{2}=\rho +3\left( \alpha H+\beta H^{2}\right), \label{eq:fried3}
\end{equation}%
similarly to the previous case, the self-accelerating solution reads, $H=\alpha / \left( 1-\beta \right)$. From this last result we can infer that the condition, $\beta < 1$, is necessary in order to have expanding solutions. In terms of the redshift the Hubble parameter turns out to be
\begin{equation}
H\left( z\right) =\frac{\alpha }{2\left( 1-\beta \right) }\left( 1+\sqrt{1+%
\frac{4\left( 1-\beta \right) \rho \left( 0\right) }{3\alpha ^{2}}\left(
1+z\right) ^{3\left( 1+\omega \right) }}\right), \label{eq:h2}
\end{equation}
where we have considered Eq. (\ref{eq:fried3}) and $\rho \left( z\right) = \rho \left( 0\right) \left( 1+z\right) ^{3\left( 1+\omega \right)}$ since non interacting fluids are under study as in the previous section. Note that in this case the Hubble parameter tends to a constant as the universe evolves, $H\left( z\rightarrow -1\right) \rightarrow \alpha /(1-\beta)$ for $\omega > -1$ and at early times $H(z\rightarrow \infty)\rightarrow \infty$; this behavior for the Hubble parameter resembles the $\Lambda$CDM model. The time derivative of Eq. (\ref{eq:ans2}) can be penned in terms of the deceleration parameter, yielding
\begin{equation}
\dot{\rho}_{de}=-3\left( \alpha +2\beta H\right) H^{2}\left( 1+q\right).
\end{equation}
On the other hand, using the equations (\ref{eq:accel}), (\ref{eq:cons1}) and (\ref{eq:cons2}) together with the above expression for $\dot{\rho}_{de}$ and the Friedmann constraint (\ref{eq:fried1}), we can write  
\begin{equation}
2\dot{H}= -p-p_{de}-3H^{2} = -\left[ (1+\omega)\rho +\left(
\alpha +2\beta H\right) H\left( 1+q\right) \right],    
\end{equation}
therefore
\begin{equation}
   q=-1+\frac{3}{2}(1+\omega)\left\lbrace \frac{1-\left( \beta +\alpha/H\right) }{1-\left( \beta +\alpha /2H\right) }\right\rbrace. 
\end{equation}
Using the conservation equation for $\rho_{de}$ and the results shown above, we can write as in the previous Ansatz a relationship between both parameters state
\begin{equation}
\omega_{de}=-1+(1+\omega)\left( \frac{\beta +\alpha /2H}{\beta +\alpha /H}\right)
\left\lbrace \frac{1-\left( \beta +\alpha/H\right) }{1-\left( \beta +\alpha /2H\right) }\right\rbrace
\end{equation}%
thus it is straightforward to verify the following conditions,
\begin{equation}
\omega_{de}\left( z\rightarrow \infty \right) \rightarrow \omega \ \ \mbox{and}%
 \ \ \omega _{de}\left( z\rightarrow -1\right) \rightarrow -1, 
\end{equation}
where $H$ is given in Eq. (\ref{eq:h2}). To sum up, the behavior of $\rho_{de}$ at early times will be dictated by the value of $\omega$ and at late times behaves as a cosmological constant independently of the value $\omega$.

\section{Allowing singularities: Phantom cosmology}
\label{sec:iv}
The previous examples of dynamical dark energy models are of cosmological interest since they have signs of the $\Lambda$CDM model at early and late times. However, the transition to a phantom regime, $\omega_{de} < -1$, is not allowed in the discussed models; the existence of such regime for dynamical dark energy models is not discarded at all by latest observations results, see for instance Ref. \cite{planck}. In order to have access to the phantom scenario, we now come up with a proposal for the dark energy density in which we consider the addition of the first derivative of the Hubble parameter to the ghost dark energy given in Eq. (\ref{eq:fried1})
\begin{equation}
   \rho_{de}= 3\left( \alpha H+\beta \dot{H}\right) = 3H\left[\alpha -\beta H (1+q) \right], 
   \label{eq:granda}
\end{equation}
with positive parameters $\alpha$, $\beta$. Notice that the dark energy proposal (\ref{eq:granda}) bears resemblance to the Granda-Oliveros (GO) holographic cutoff for dark energy which is written as, $\rho_{GO} = 3(\gamma H^{2}+\delta \dot{H})$, being $\gamma, \delta$ arbitrary parameters \cite{grandaoliveros}. It is worthy to mention that in order to have phantom scenario in FLRW cosmologies, generalized forms (including powers or derivatives of the Hubble parameter) for the energy density of dark energy have been studied extensively, see for instance \cite{inhomogeneous}. We must have in mind that the GO cutoff is simply a generalization of the holographic model given as, $\rho_{de} \propto R$, where $R$ is the Ricci scalar, which in turn is written as, $6(2H^{2}+\dot{H})$, for a flat FLRW background, see Ref. \cite{ricci}.

Considering the Friedmann constraint (\ref{eq:fried1}) with $\rho=0$ and the above energy density (\ref{eq:granda}) for dark energy, we can solve the resulting first order differential equation for the Hubble parameter, yielding
\begin{equation}
   H(t) = \frac{\alpha}{1+(\alpha/H_{0} - 1)\exp \left[ \left( \alpha /\beta \right) \left( t-t_{0}\right) \right]}, \label{eq:sing1}
\end{equation}
where $H_{0}$ is the Hubble constant defined from the initial condition $H(t=t_{0}) = H_{0}$. For $\alpha = H_{0}$, we recover a de Sitter evolution, i.e, the Hubble parameter (\ref{eq:sing1}) becomes the constant $\alpha$. Integrating the above result we can obtain the scale factor explicitly as a function of cosmic time, therefore
\begin{equation}
a(t) =a_{0}\exp \left[\alpha \left( t-t_{0}\right) + \ln \left(\frac{\alpha /H_{0}}{1+\left( \alpha /H_{0}-1\right) \exp \left[\left(\alpha /\beta \right) \left( t-t_{0}\right) \right]}\right)^{\beta}\right], \label{eq:scale2}
\end{equation}
being $a_{0}$ a constant defined as $a(t=t_{0}) = a_{0}$. It is worthy to mention that the Hubble parameter (\ref{eq:sing1}) becomes singular at $t=t_{s}$, where
\begin{equation}
t_{s} = t_{0}+\frac{\beta }{\alpha }\ln \left( \frac{1}{1-\alpha /H_{0}}\right),
\end{equation}
and $\alpha/H_{0} < 1$. Notice that under this last condition the scale factor (\ref{eq:scale2}) remains bounded, then is its first derivative of $a(t)$ that diverges at $t_{s}$. This kind of behavior is characteristic of a Type III singularity. We would like to mention that in the GO model a future singularity of Type I (Big Rip) is allowed for $\gamma < 1$ since in this case we can write 
\begin{equation}
    H(t) = \frac{H_{0}}{1-H_{0}[(1-\gamma)/\delta](t-t_{0})} \ \ \mbox{and} \ \  t_{s} =t_{0}+\left(\frac{\delta}{1-\gamma} \right)H^{-1}_{0}.
\end{equation}
Then the future evolution of these dark energy models differs considerably despite the similarities between both models; the full classification for future singularities can be seen in Ref. \cite{classi}. From Eqs. (\ref{eq:cons1}), (\ref{eq:granda}) and (\ref{eq:sing1}), we can obtain by direct calculation the following expression
\begin{equation}
    \omega_{de}(t) = -1 + \frac{2}{3}\frac{(\alpha/H_{0}-1)}{\beta}\exp\left[\frac{\alpha}{\beta}(t-t_{0})\right],
    \label{eq:sfw}
\end{equation}
therefore this parameter state describes a phantom fluid under the assumption, $\alpha/H_{0} < 1$ and takes a finite value for $t=t_{s}$.

On the other hand, for $\alpha /H_{0}>1$ we can verify the following conditions for (\ref{eq:sing1})
\begin{align}
& H\left(t\ll t_{0}\right) \rightarrow \alpha,\\
& H\left(t\gg t_{0}\right) \sim \left(\frac{\alpha 
}{\alpha /H_{0}-1}\right)\exp \left[-\left( \alpha /\beta \right) \left(t-t_{0}\right)\right],
\end{align}
i.e., this universe evolves from a de Sitter-like expansion at early times to a final stage given by $H\left(t\rightarrow \infty \right) \longrightarrow 0$.  

\subsection{Scalar field correspondence}
\label{sec:sf}
The pressure and energy densities associated to a scalar field, $\phi$, are given as \cite{faraoni}
\begin{equation}
    p_{\phi} = \frac{\dot{\phi}^{2}}{2}-V(\phi), \ \ \ \ \rho_{\phi} = \frac{\dot{\phi}^{2}}{2}+V(\phi),
\end{equation}
for a flat FLRW geometry, being $V(\phi)$ the scalar field potential. Then, for a barotropic equation of state we can write the parameter state as follows
\begin{equation}
    \omega_{\phi} = \frac{\dot{\phi}^{2}-2V(\phi)}{\dot{\phi}^{2}+2V(\phi)}.
\end{equation}
Comparing the last expression with Eq. (\ref{eq:sfw}) one gets
\begin{equation}
    \frac{\dot{\phi}^{2}-2V(\phi)}{\dot{\phi}^{2}+2V(\phi)} = -1 + \frac{2}{3}\frac{(\alpha/H_{0}-1)}{\beta}\exp\left[\frac{\alpha}{\beta}(t-t_{0})\right], 
\end{equation}
together with the equation 
\begin{equation}
    \rho_{\phi} = \frac{\dot{\phi}^{2}}{2}+V(\phi) = 3\left( \alpha H+\beta \dot{H}\right),
\end{equation}
and the Hubble parameter (\ref{eq:sing1}), we can solve to obtain explicit results for the scalar field and the potential, to wit
\begin{equation}
    \phi(t) = \phi_{0} + \sqrt{\frac{2\beta}{\alpha/H_{0}-1}}\ln \left[\frac{H_{0}}{\alpha} \left\lbrace 1 + \left(\frac{\alpha}{H_{0}} - 1\right) \exp \left[\frac{\alpha}{2\beta}(t-t_{0}) \right]\right\rbrace \right],
\end{equation}
where $\phi_{0}$ is an integration constant given by $\phi(t = t_{0}) = \phi_{0}$, besides 
\begin{equation}
    V(\phi) = \frac{\alpha^{2}}{\beta}\frac{ \left\lbrace 3\beta - (\alpha/H_{0}-1)^{-1}\left(1 - (\alpha/H_{0})\exp \left[\sqrt{(\alpha/H_{0}-1)/2\beta}(\phi-\phi_{0}) \right]\right)^{2}\right\rbrace }{\left\lbrace 1 + (\alpha/H_{0}-1)^{-1}\left(1 - (\alpha/H_{0})\exp \left[\sqrt{(\alpha/H_{0}-1)/2\beta}(\phi-\phi_{0}) \right]\right)^{2}\right\rbrace^{2}}.
\end{equation}
The exponential potential has been widely explored in the scalar field approach, it is well known that such potential can produce accelerated cosmic expansion. However, as can be seen in our results, the condition $\alpha/H_{0} > 1$ must be guaranteed in order to have well defined square roots; thus if we carry the aforementioned condition to Eq. (\ref{eq:sfw}), we find that this scalar field describes a quintessence scenario despite the correspondence established with a phantom model. As found in Ref. \cite{vikman}, the dynamical transition to a phantom evolution by a standard scalar field is in general physically implausible. 

\subsection{A novel proposal for dark energy}
\label{sec:v}
In this section we propose a dark energy density that depends only on $H$ as in the examples reviewed in Section \ref{sec:ii}. We consider the following Ansatz for the energy density
\begin{equation}
\rho_{de}=3\alpha^{2}H^{4}, \label{eq:friednovel}
\end{equation}
this specific dependence on $H$ is encouraged from two sources, the Einstein-Gauss-Bonnet gravity in five dimensions, where the Friedmann constraint involves terms of this type \cite{salgado1} and from the introduction of correction terms of this form in the inflationary epoch, which are interpreted as entropic forces \cite{entropic}. The Friedmann constraint (\ref{eq:fried1}) in this case reads 
\begin{equation}
3H^{2}=\rho +3\alpha ^{2}H^{4},
\end{equation}
from the previous equation a self-accelerated solution appears again for $\rho=0$ given as $H=1/\alpha$. For $\rho \neq 0$ the solution for the Hubble parameter is obtained from a quartic algebraic equation and is given by
\begin{equation}
H_{\pm }^{2}\left( z\right) = H_{\pm }^{2}\left( z_{s}\right) \left( 1\pm 
\sqrt{1-\frac{\rho \left( z\right) }{\rho \left( z_{z}\right) }}\right),
\end{equation}
where we have defined $H_{\pm}\left(z_{s}\right) := 1/\sqrt{2}\alpha$ and $\rho \left(
z_{z}\right) =3/4\alpha ^{2}$. In order to avoid a complex nature in the Hubble parameter we observe that the dark matter density $\rho(z)$ has an upper bound given by $\rho(z_{z})$. Using this result for $H$ we can determine the following expression for the dark energy density
\begin{equation}
\rho_{de}\left( z\right) = \rho \left( z_{z}\right) \left(1\pm \sqrt{1-\frac{\rho \left( z\right) }{\rho \left( z_{z}\right) }}\right)^{2}.\label{eq:dens}
\end{equation}
Using the Friedmann constraint (\ref{eq:friednovel}), it is straightforward to obtain the normalization condition for the fractional energy density parameters as performed previously, $\Omega_{i} = \rho_{i}/3H^{2}$, then the value of the parameter $\alpha$ can be obtained by evaluating our expression at present time ($z=0$)
\begin{equation}
1 = \Omega \left(0\right) +\alpha ^{2}H^{2}\left( 0\right) \rightarrow
\alpha ^{2}=\frac{1-\Omega \left( 0\right) }{H^{2}\left( 0\right) }, 
\end{equation}
when $\rho(z) = \rho(z_{z})$ in Eq. (\ref{eq:dens}), we obtain the equality $\rho_{de}\left( z\right) = \rho \left( z_{z}\right)$, therefore
\begin{equation}
\rho_{de}\left( z_{z}\right) =\rho \left( z_{z}\right) =\frac{3H^{2}\left(
0\right) }{4\left[ 1-\Omega \left( 0\right) \right] },  \tag{45}
\end{equation}%
if we define, $\lambda :=4\left[ 1-\Omega \left( 0\right) \right] \Omega \left( 0\right)$, we can write for the Hubble parameter
\begin{equation}
H_{+}^{2}\left( z\right) = \left( \frac{1+\sqrt{1-\lambda \left( 1+z\right)^{3}}}{2\left[ 1-\Omega \left( 0\right) \right] }\right) H^{2}\left(0\right), 
\end{equation}
where we have considered the solution coming from (\ref{eq:cons2}) for dark matter sector with $\omega=0$, as in the previous section, we are dealing with no interacting fluids. Then one gets for the dark energy density
\begin{equation}
\rho_{de}\left( z\right) = 3\left[ 1-\Omega \left( 0\right) \right] \left(\frac{1+\sqrt{1-\lambda \left( 1+z\right)^{3}}}{2\left[ 1-\Omega \left(0\right) \right]}\right) ^{2}H^{2}\left( 0\right). \label{eq:c1}
\end{equation}
The cosmic coincidence parameter, $r(z)$, can be constructed by means of the quotient between the energy densities of both components, $r(z) := \rho(z)/\rho_{de}(z)$, which in turn results as
\begin{equation}
r\left(z\right) =\frac{\lambda \left( 1+z\right) ^{3}}{\left( 1+\sqrt{1-\lambda \left( 1+z\right)^{3}}\right)^{2}}, 
\end{equation}%
this expression of $r$ is useful if we desire to fix the value of $\lambda$. In Ref. \cite{trans} was established that independently of the cosmological model, the universe has a transition from decelerated to accelerated expansion approximately at $z = 0.64$, this means that around that redshift value the dark energy dominance began, therefore $\rho_{de} \approx \rho$; this condition leads to $\lambda \approx 0.23$ if we consider the aforementioned value for the redshift, note that $r\left( z\rightarrow -1\right) \rightarrow 0$, which is consistent with a growing dark energy model. As discussed previously, from the Eq. (\ref{eq:cons1}) and our Ansatz for $\rho_{de}$ we can write
\begin{equation}
1+q =\frac{3}{4}\left(1+\omega_{de}\right), 
\end{equation}
and from the time derivative of the Friedmann constraint we have for the dark matter sector
\begin{equation}
1+q=\frac{3}{2}\left(1+\omega \right) \frac{r}{r-1}.
\end{equation}
From these latter results we can write the dark energy parameter state as a function of the the coincidence parameter and the parameter state $\omega$, yielding
\begin{equation}
\omega_{de}(z) = -1 - 2\left(1+\omega \right)\frac{r(z)}{1-r(z)}. \label{eq:c2}
\end{equation}
Here lies the singular nature of this cosmological model. At the moment at which the dark energy domination begins, $r \approx 1$, its parameter state diverges negatively and eventually evolves to a cosmological constant like behavior since $r\left( z\rightarrow -1\right) \rightarrow 0$, i.e., the phantom behavior in this scenario is transitory; this is not atypical in cosmology, as pointed out in Ref. \cite{transient}, the consideration of some mechanisms could help to prevent (or to kick away) singularities, then the phantom stage can be seen as a transient epoch in the cosmic evolution. See also Ref. \cite{transient2}, where the consideration of quantum gravity effects in a phantom scalar-tensor model leads to a future singularity-free cosmic evolution. Notice that in our model we do not depend on extra instruments to get over the phantom stage, we only reckon on the fact that dark energy evolves displaying a growing behavior. According to our results, this dark energy model is applicable only from the redshift value at which $\rho_{de}$ begins to dominate until the far future, $z=-1$. 

\begin{figure}[htbp!]
\centering
\includegraphics[scale=0.75]{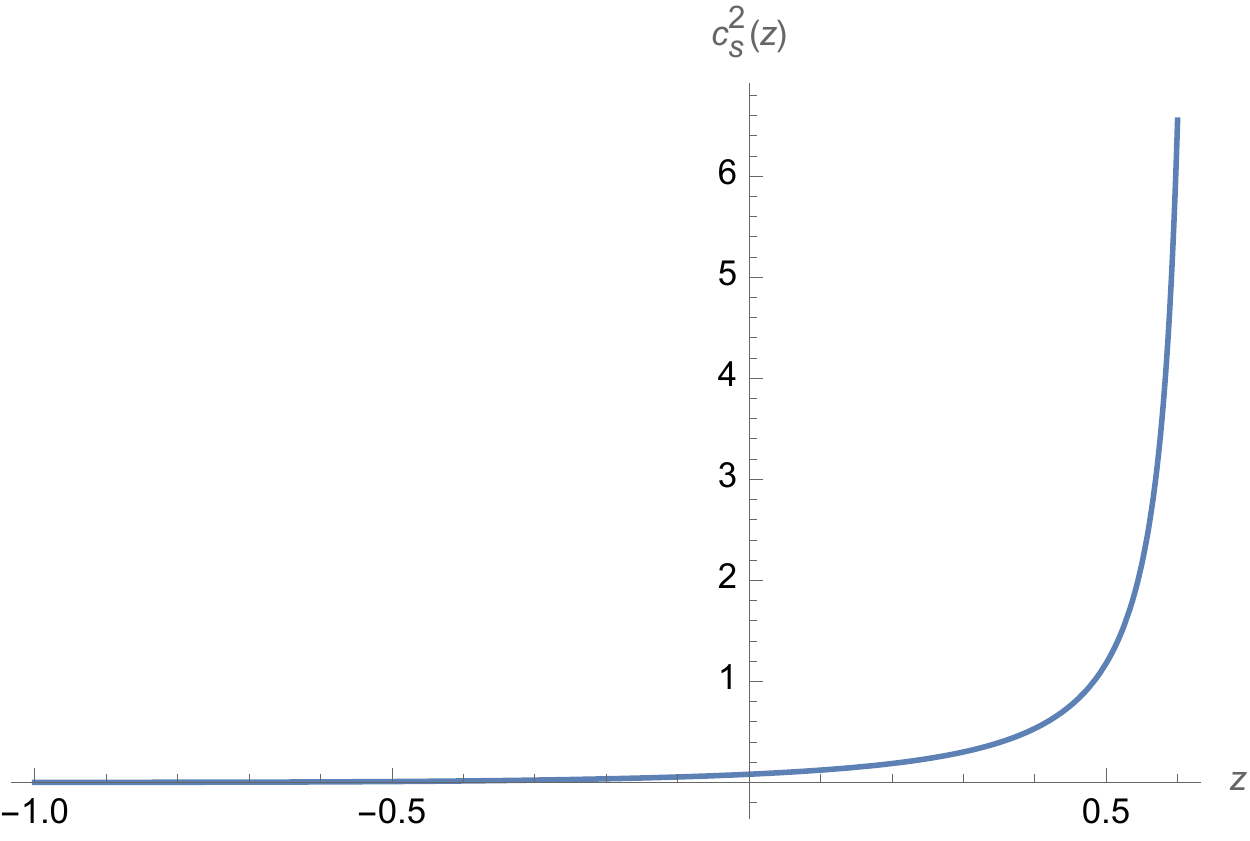}
\caption{Behavior of $c^{2}_{s}$. For this plot we have considered the expressions (\ref{eq:c1}) and (\ref{eq:c2}) with the value 0.23 for the parameter $\lambda$ and the matter sector described as dark matter, $\omega=0$.}
\label{fig:adiabatic}
\end{figure}

For this model the squared adiabatic sound speed has the usual form
\begin{equation}
c^{2}_{s} = \frac{\dot{p}_{de}}{\dot{\rho}_{de}} = \omega_{de} + \dot{\omega}_{de}\frac{\rho_{de}}{\dot{\rho}_{de}}, 
\end{equation}
where a barotropic equation of state was considered. In Fig. (\ref{fig:adiabatic}) we show the behavior of this quantity in terms of the redshift and within the region of validity for the cosmological model. As can be seen we have, $c^{2}_{s} > 0$, this is signal of stability in this dark energy model. In the case of the dark energy model (\ref{eq:granda}) we have unstable behavior, i.e., $c^{2}_{s} < 0$, for $\alpha < \beta$ and $\alpha > \beta$. However, we must have in mind that our proposal extends the ghost dark energy model \cite{ansatz1} by the addition of the $\dot{H}$ term.     

\section{Concluding remarks}
\label{sec:vi}
We explored some dynamical dark energy models emerging from the QCD Veneziano ghost. We enlarge the cosmological description of this kind of model by discussing some extensions of the ghost dark energy since it has been insufficiently treated in the literature. For the original model as well as for its first extension (given by the addition of a quadratic term of the Hubble parameter) we find that at early and late times the behavior of the dark energy component is prescribed by the dark matter sector, i.e., the parameter state of the components are related to each other, although we are not considering interacting fluids. In summary, for both cosmological scenarios the cosmic evolution tends to a de Sitter expansion, thus the dynamical transition to a phantom evolution is not allowed. 

Mimicking the holographic approach for dark energy, we have considered the inclusion of the $\dot{H}$ term in the energy density of the original ghost dark energy. In this case we found that the model allows a future singularity of Type III, or in other words, the scale factor remains bounded despite the singular fate of the universe. The correspondence between this model and the standard scalar field approach was explored, obtaining that the scalar field describes at most a quintessence scenario with a specific form for the potential, $V(\phi)$, which is reconstructed from the background dynamics. An interesting extension for this scalar field correspondence could be given by establishing a connection with other generalized models for the scalar field, where the phantom stage is allowed. For instance, in Ref. \cite{eoin} was found that the Horndeski model is compatible with a phantom cosmology and alleviates the $H_{0}$ and $\sigma_{8}$ tensions. We will explore this elsewhere.  

Additionally, inspired from other cosmological scenarios, we have also discussed our fresh proposal for dark energy which can be seen as a quartic version of the original ghost dark energy model. As discussed previously, our scheme admits a phantom scenario whose singularity lies only on the parameter state of dark energy and appears in the deceleration-acceleration transition stage; from there the model tends to a de Sitter evolution, i.e., our approach allows a transient phantom regime. A signal of stability for this dark energy model is obtained from the positivity of the squared adiabatic sound speed. 

On the other hand, the extended model depending on $\dot{H}$ is unstable under the squared adiabatic sound speed criterion. However, we leave for future investigation if these extensions of the ghost dark energy are also a ghost model. In such case the models describe non physical degrees of freedom for which the signal of instability is irrelevant and is consequence of treating them as conventional propagating degrees of freedom satisfying a classical equation of motion. Finally, we would like to comment that the quartic model for dark energy can be also extended by considering the inclusion of a $\dot{H}$ term in the energy density; in such case depending on the values of the cosmological parameters, the model preserves its phantom nature or could also provide a quintessence scenario. We will review this subject elsewhere.  

\section*{Acknowledgments}
MC work has been supported by S.N.I. (CONACyT-M\'exico). SL appreciates the interesting comments made by R. Labarca (USACH). GS acknowledges encouragement from Instituto de F\'\i sica, Pontificia Universidad Cat\'olica de Valpara\'\i so, Chile. MC dedicates this article to the memory of Dr. Riccardo Capovilla, mentor, colleague and friend.

\end{document}